\documentclass[conference]{IEEEtran}
\usepackage[right=0.64in, bottom=1in, top=0.7in, left=0.63in]{geometry}

\IEEEoverridecommandlockouts
\usepackage{listings}
\usepackage{graphicx}
\usepackage{caption}
\usepackage{subcaption}
\usepackage{todonotes}
\usepackage{paralist}
\usepackage{makecell}
\usepackage{url}
\usepackage{svg}

\usepackage[compact]{titlesec}
\titlespacing{\section}{0pt}{1ex}{0ex}
\titlespacing{\subsection}{0pt}{0.1ex}{0ex}\titlespacing{\subsubsection}{0pt}{0.1ex}{0ex}    

\lstdefinelanguage{json}{
    basicstyle=\normalfont\ttfamily,
    numbers=right,
    numberstyle=\scriptsize,
    stepnumber=1,
    numbersep=8pt,
    showstringspaces=false,
    breaklines=true,
    frame=lines,
    basicstyle=\ttfamily\footnotesize,
  morestring=[b]",
  moredelim=[s][\bfseries\color{blue}]{<}{\ },
  moredelim=[s][\bfseries\color{blue}]{</}{>},
  morecomment=[s]{--}{'},
  commentstyle=\color{red},
  stringstyle=\color{blue},
  identifierstyle=\color{black}
}

\begin{document}
%
\title{Towards Incident Response Orchestration and Automation for the Advanced Metering Infrastructure}

\author{\\[-9ex]\IEEEauthorblockN{Alexios Lekidis\IEEEauthorrefmark{1}, Vasileios Mavroeidis\IEEEauthorrefmark{2} and Konstantinos Fysarakis\IEEEauthorrefmark{3},
\IEEEoverridecommandlockouts }

\IEEEauthorrefmark{1} University of Thessaly, Greece, 
\IEEEauthorrefmark{2} University of Oslo, Norway \\
\IEEEauthorrefmark{3} Sphynx Analytics Limited, Cyprus 
\footnotesize{Email:\IEEEauthorrefmark{1}alekidis@uth.gr,\IEEEauthorrefmark{2}vasileim@ifi.uio.no, \IEEEauthorrefmark{3}fysarakis@sphynx.ch}}


\maketitle

\begin{abstract}
The threat landscape of industrial infrastructures has expanded exponentially over the last few years. Such infrastructures include services such as the smart meter data exchange that should have real-time availability. Smart meters constitute the main component of the Advanced Metering Infrastructure, and their measurements are also used as historical data for forecasting the energy demand to avoid load peaks that could lead to blackouts within specific areas. Hence, a comprehensive Incident Response plan must be in place to ensure high service availability in case of cyber-attacks or operational errors. Currently, utility operators execute such plans mostly manually, requiring extensive time, effort, and domain expertise, and they are prone to human errors. In this paper, we present a method to provide an orchestrated and highly automated Incident Response plan targeting specific use cases and attack scenarios in the energy sector, including steps for preparedness, detection and analysis, containment, eradication, recovery, and post-incident activity through the use of playbooks. In particular, we use the OASIS Collaborative Automated Course of Action Operations (CACAO) standard to define highly automatable workflows in support of cyber security operations for the Advanced Metering Infrastructure. The proposed method is validated through an Advanced Metering Infrastructure testbed where the most prominent cyber-attacks are emulated, and playbooks are instantiated to ensure rapid response for the containment and eradication of the threat, business continuity on the smart meter data exchange service, and compliance with incident reporting requirements.

\end{abstract}

\section{Introduction} \label{sec:intro}

Cyber-attacks against Operators of Essential Services (OES) have increased over the recent years, and utility stakeholders, specifically, identify sophisticated attacks as a top challenge \cite{threatLandscape}.

Nevertheless, conducted studies \cite{shaked2023operations} have shown that 35\% of utilities have no response plan in place. Moreover, utilities with a response plan mainly focus on manual actions performed by operators and IT engineers to restore the systems to normal operations and eradicate the threat. Specifically, for Advanced Metering Infrastructure (AMI) systems, when smart meter data cannot be transmitted by the smart meters due to networking, operational, or even cyber security issues, utilities use traditional metering measures with on-site operators to make the data available in the AMI Headend. This is also a time-consuming process and is performed based on the operators' availability. 

However, the unavailability of smart meter measurements may cause cascading effects on more critical utility services, such as forecasting electricity consumption, which should be accommodated through sufficient production sources to avoid load demand peaks and blackouts. The situation is further complicated by the strict incident reporting requirements that the NIS2 Directive \cite{nis2parliament} imposes on utilities (i.e., essential and important entities). 


Addressing the above aspects is a challenge for utilities, requiring the development of new services and enhancing existing ones with automation. Motivated by this landscape, PHOENI2X\footnote{https://phoeni2x.eu/} is a Horizon Europe project that aims to address this challenge by providing tools and mechanisms as essential enablers to ensure cyber resilience, adopting a recently published conceptual blueprint in support of architecting and establishing interoperable Cyber Security Operations Centres that combine capacity for Shared Situational Awareness, Coordinated Response, and Joint Preparedness \cite{Blueprint}. Building upon that, this paper presents a method based on Incident Response (IR) playbooks to orchestrate and automate, to the extent possible, the IR phases regarding AMI environments, leveraging a set of relevant technological building blocks offered by PHOENI2X. By introducing an IR plan and automating parts of the process, a utility can ensure the highest availability of its services, cyber resilience, and business continuity through the prevention, detection, mitigation, and recovery from incidents.

The concrete contributions of this paper are:
\begin{compactitem}
    \item Presentation and application of a method for automating and orchestrating incident response in the light of cyber-attacks against AMI systems using CACAO playbooks, as leveraged in PHOENI2X. 
    \item Utilization and demonstration of the CACAO security playbooks standard (a horizontal approach) on specific use cases in the energy sector, validating its adequacy to satisfy any sector-specific peculiarities in IR while also emphasizing the importance of exchanging such interoperable playbooks for collaborative defense.
    \item Evaluation of the proposed method in a realistic emulated AMI testbed where False Data Injection (FDI) and Denial of Service (DoS) attacks are performed, detected, and mitigated through custom-tailored playbooks.
\end{compactitem}

The rest of the paper is organized as follows. Section \ref{sec:background} provides an overview of the AMI (eco)system and the use of playbooks for incident response automation. Section \ref{sec:method} presents the proposed method by focusing on the IR requirements and response automation mechanisms tailored to AMI systems. Then, Section \ref{sec:caseStudy} demonstrates an application of the proposed method in an AMI testbed where emulated cyber-attacks are performed, and playbooks are instantiated to automate the incident response process. A discussion of evaluation metrics for the proposed method, ongoing work, and future directions follows in Section \ref{sec:discussion}. Finally, Section \ref{sec:conclusion} provides the concluding remarks.

\section{background} \label{sec:background}

\subsection{Advanced Metering Infrastructure}

\begin{figure}[bthp!]
         \includegraphics[width=1\columnwidth]{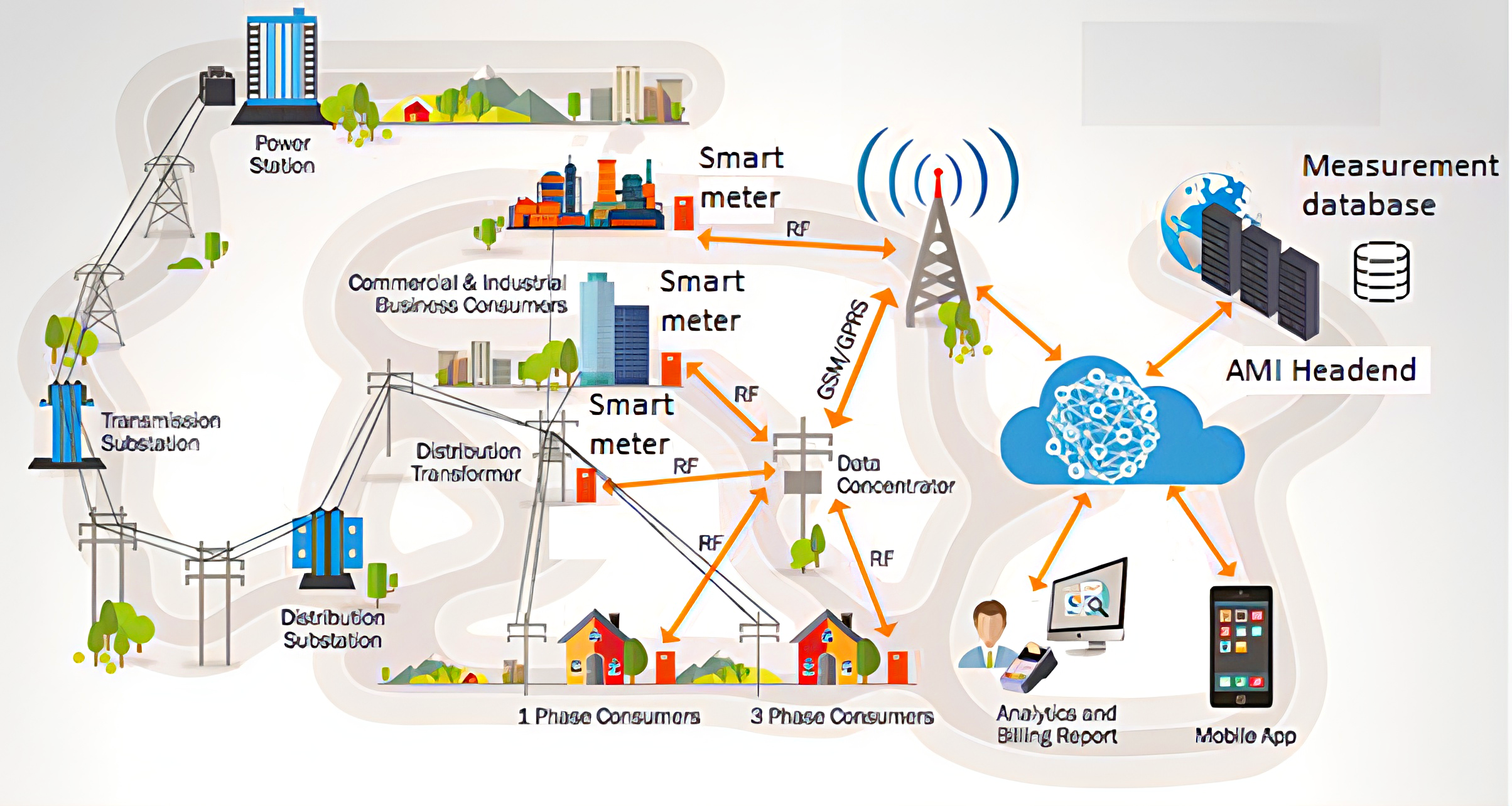}
     \caption{Advanced Metering Infrastructure communication}
     \vspace{-3mm}
     \label{fig:amiOverview}
\end{figure}

Smart meters allow for the real-time availability of electricity production and consumption data (i.e., current, voltage, active, reactive, and apparent power) as well as the implementation of automated customer billing scenarios based on their overall consumption, which is achieved through the AMI. The ecosystem (depicted in Fig. \ref{fig:amiOverview}) includes smart meters deployed in consumer households, near distribution transformers, near commercial and industrial business consumers (e.g., manufacturing systems providers, hospitals, and banks) as well as a central platform for collecting the measurements, called AMI Headend \cite{sarfi2012ami}. The Headend is a dedicated software installed on a server to collect smart meter data from different areas and perform validation and pre-processing. It is also used to perform configuration setups, modifications, software updates, and ad-hoc requests to the smart meters.

The smart meters use cellular connectivity, such as GPRS or 3G, to transmit the electricity consumption data to the AMI Headend. The Headend includes a database for storing and aggregating the received data. The data are also available in analytics and billing reports in web-based platforms and the mobile applications of the individual consumers (Fig. \ref{fig:amiOverview}). 

There are multiple protocols involved in AMI communications. DLMS/COSEM is a widely used communication protocol for metering devices \cite{butun2020security}. DLMS stands for "Device Language Message Specification" and describes a concept for modeling communication entities. COSEM stands for "COmpanion Specification for Energy Metering" and sets the rules for data exchange with energy meters. Specifically, in existing utility business processes, smart meter measurements are transmitted from the locations depicted in Fig. \ref{fig:amiOverview} using modules for DLMS/COSEM over GPRS/3G communication\footnote{https://www.landisgyr.eu/webfoo/wp-content/uploads/2012/09/LandisGyr-HES\_Product-Description.pdf}.

\subsection{Incident Response Playbooks and Automation}
The cyber security industry has recently seen a proliferation of Security Orchestration, Automation and Response (SOAR)\footnote{https://www.gartner.com/en/documents/3990720} and Extended Detection and Response (XDR)\footnote{https://www.gartner.com/en/documents/4007995} offerings that utilize playbooks to orchestrate and automate cyber security operations. Such solutions also offer the ability to represent playbooks graphically, typically in the form of sequence diagrams, and edit and fine-tune them in a user-friendly manner with minimal coding.

In their basic form, security playbooks are documents that present formal processes to be followed in the context of different cyber security incidents and operations, including threat detection and response. Multiple resources provide reference playbooks and playbook guides, such as those made available by NIST \cite{souppaya2017guide}, SANS\footnote{https://www.sans.org/presentations/ir-playbooks/}, security agencies\footnote{https://github.com/cisagov/shareable-soar-workflows} and others\footnote{https://www.incidentresponse.com/playbooks/}. 

While the number of products and services that utilize machine-readable and executable playbooks is constantly increasing, these typically rely on proprietary formats and are tightly integrated within the SOAR and XDR systems that utilize them\footnote{https://docs.fortinet.com/document/fortisoar/6.0.0/playbooks-guide}. A limitation of this approach is the non-interoperable nature of these playbooks, resulting in most cases in vendor lock-in or, for the brave and well-resourced entities, a very cumbersome and expensive migration to a new SOAR where all the existing playbooks would have to be converted and tuned based on the SOAR's underlying supported schema and encoding. Similarly, a non-standardized format for playbooks thwarts their potential of being exchanged in support of collective defense, a paradigm that the cyber security community has, to a certain extent, been very successful in implementing for the exchange of CTI. 

Consequently, there is a pressing need for publicly available, automatable, extendable, and shareable playbooks. This motivated the standards-developing organization OASIS\footnote{https://www.oasis-open.org/org/} and, in particular, a formulated technical committee with like-minded individuals to develop the CACAO standard\footnote{https://docs.oasis-open.org/cacao/security-playbooks/v2.0/security-playbooks-v2.0.html} aiming to establish a common schema for compiling security playbooks able to be shared across various sectors and organizations, in a vendor-agnostic manner. In addition, works on the advantages of coupling and exchanging playbooks with Cyber Threat Intelligence (CTI) and prototypical implementations have emerged \cite{mavroeidis2021integration, mavroeidis2022cybersecurity}.


Overall, while playbooks are gaining popularity, the cyber security IR landscape is currently dominated by proprietary playbook specifications or non-cybersecurity relevant open specifications (e.g., BPMN notation \cite{BPMN_CACAO}). The inclusion of machine-speed threat situational awareness exchange mechanisms within the playbooks and the ability to encode, share, and automate the actuation of IR tradecraft, as those enabled by CACAO, including reporting (generating reports aligned with OES requirements), is a pressing need in this regard.

Recent efforts have introduced playbooks for Industrial Control Systems, which are basic blocks of OES infrastructures. Specifically, the authors in \cite{mccarty2023cybersecurity} demonstrated a SOAR for enhancing the resilience of a wind energy environment. Further work has focused on artificial intelligence to provide automation in the creation of IR workflows. In \cite{allison2023digital}, the authors demonstrated a Digital-Twin-based approach for handling incidents in OES infrastructures using Programmable Logic Controllers (PLCs). Furthermore, the concept of autonomous attack mitigation is also investigated in \cite{mern2022autonomous} through a deep reinforcement learning approach, which builds a model based on PLCs. Finally, the authors in \cite{empl2023generating} used ICS security advisories to automatically generate CACAO playbooks with remediation steps pertaining to vulnerability management.

Even though the use of IR playbooks is increasing in literature, to the best of our knowledge, there is still limited work on orchestrating/automating the entire IR lifecycle, including the preparation, cyber-attack detection, and post-incident reporting steps, considering the recent (EU) regulatory requirements. This work, to a certain extent, focuses on tackling this challenge through the use of CACAO-based security playbooks.

\section{Responding to attacks on the AMI} \label{sec:method}
\subsection{Threats \& Incident Response Requirements} \label{sec:Reqs}
To kick-start our research, a risk assessment on AMI was conducted with a focus on business continuity. The AMI system was divided into its main services linked with the transmission of measurements from the smart meters and, in particular, customer households, near distribution transformers, and near commercial and industrial business consumers (i.e., OES with critical buildings and high-power consumption). 

This assessment identified different threat scenarios against AMI systems \cite{li2019review} where detection, mitigation, and other IR processes are needed. Furthermore, we have concluded that even though there are high-impact scenarios for the AMI system, such as malware spreading attacks \cite{zhang2020modeling}, or network DoS \cite{huseinovic2020survey}, they are not very likely to manifest in large-scale AMI systems. The reason is that energy utilities usually have multiple protection schemes against them. Instead, smart meter data tampering or electricity theft scenarios have a higher likelihood of occurrence in the AMI system. However, their impact is not high for the operation of the electricity infrastructure. Still, they can lead to supply chain incidents pertaining to different functions of the OES handled by partner organizations, such as billing. For example, data tampering attacks have occurred in the past in the smart meters of Electric Vehicle (EV) charging stations owned by a company in the Public Power Corporation (PPC) Group, causing each customer to be charged 2000 dollars on average\footnote{https://medium.com/@ekarabatsakis/how-customers-helped-us-survive-a-cyber-attack-5ddd2729c3f3}.

Hence, for this study, the most prominent attacks relevant to AMI are False Data Injection (FDI) and alternation or loss of smart meter data measurements. In both scenarios, to achieve the desired goal, an attack has to be executed continuously for many consecutive measurements that will be exchanged and not just occasionally. Moreover, except for the impacts mentioned above, such alternation would lead to inaccurate AMI measurements transmitted to the utility's control center and differences in the amount of energy forecasted to be produced by the utility's power plants, as the historical measurements from the AMI system are incorrect.

For this research, to establish and implement a tailored IR plan, we focused on the following steps:

\begin{compactenum}
    \item \textbf{Preparation}: The utility has to define an IR team with clear roles, responsibilities, and security processes, determine communication channels, contact points, incident reporting requirements and mechanisms, deploy security controls and tools for detection, analysis, and incident management, have in place incident analysis resources, and conduct user awareness training (e.g., use of security password policy) including IR training utilizing different approaches like Cyber-Range training on responding to different attack types, such as phishing and ransomware tailored to the AMI system \cite{lieskovan2022security}. Additionally, periodic risk assessments shall be performed to compute the operational/cyber-attack impact on each component of the AMI system based on its criticality and how likely the attack is to occur and further propagate in the ecosystem.
    \item \textbf{Cyber-attack detection and analysis}: Personnel should continuously monitor the received measurements in the AMI Headend to identify misalignments or deviations in certain areas. Additionally, network, host monitoring, and information and event management systems are central in the utility's control centers to detect cyber-attacks at the network or host level. These measures are used to collect evidence to investigate the root cause of anomalies, reason if they were caused due to a cyber-attack or an operational fault in the AMI system, triage, and further determine the execution of an appropriate IR process. At the same time, when plausible, forensic analysis should be performed.  
    \item \textbf{Containment}: Once the incident has been determined to be a cyber-attack, containment procedures are executed. For example, a process relevant to the scenario of a compromised AMI could require isolating the infected parts of the AMI and preventing the further propagation of the attack to other systems, for example, through lateral movement techniques. Specific containment measures can include isolating the infected devices and subsystems (e.g., in a Virtual Local Area Network network segment) and configuring firewall rules to ensure that the attacker is prohibited access to the AMI system. 
    \item \textbf{Eradication and Recovery}: Upon incident containment, the IR team proceeds with eradicating the threat and restoring the system to its full operational capacity. Eradication and recovery actions usually include removing any traces of the threat in the infrastructure, reinstalling the firmware in the smart meters, and performing a configuration reset on both the smart meters and the AMI Headend. Moreover, often in AMI systems, a backup device on hot or cold standby is also enabled to ensure the AMI system operates smoothly while the threat is eradicated. Such a device may be the AMI headend or even the smart meter (virtual smart meter). 
    \item \textbf{Incident Reporting}: In case of an incident that disrupts utility services, EU countries should inform their National Competent Authority and Computer Security Incident Response Team (CSIRT). Furthermore, the NIS2 Directive \cite{nis2parliament} states that OES are required to have a minimum baseline set of security capabilities. To incorporate and implement the NIS Directive, each EU Member State has defined reporting and information-sharing processes between the National Competent Authorities and the OES. This is usually performed through a report along with further technical and operational evidence of the incident (e.g., log files, indicators of compromise, employed attack methods). The report follows the national legislation framework and the European reporting and information-sharing requirements. In particular, as stated in NIS2 \cite{nis2parliament} \textit{where essential or important entities become aware of a significant incident, they should be required to submit an early warning without undue delay and, in any event, within 24 hours. That early warning should be followed by an incident notification. The entities concerned should submit an incident notification without undue delay and, in any event, within 72 hours of becoming aware of the significant incident, with the aim, in particular, of updating information submitted through the early warning and indicating an initial assessment of the significant incident, including its severity and impact, as well as indicators of compromise, where available. A final report should be submitted no later than one month after the incident notification.}
\end{compactenum}

\subsection{AMI-tailored Incident Response Automation Using CACAO Playbooks}
The PHOENI2X Cyber Resilience Framework (CRF) provides AI-assisted orchestration, automation and response capabilities, covering business continuity, IR, and information exchange, tailored to the needs of OES and the EU MS National Authorities entrusted with cyber security through the deployment of PHOENI2X Cyber Resilience Centres (PHOENI2X CRCs). 
Exhaustively going through the building blocks of PHOENI2X CRCs is beyond the scope of this paper; they are also detailed in the concept paper published by the project's consortium \cite{PHOENI2X_Concept}.


Nevertheless, fundamental to the PHOENI2X Resilience Orchestration and Response (ROAR) capabilities and the work presented herein are the ROAR features of the CRCs, which execute and orchestrate actions encoded within Resilience Playbooks (RPs). RPs provide a structured machine-processable sequence of actions comprising the organization's resilience, business continuity, recovery, and IR processes. Each action represents a procedure (e.g., isolating a system - containment). Thus, through RPs, organizations can specify, automate the execution (via the purpose-built execution and orchestration engine), monitor the progress, and assess the effectiveness of all their business continuity, recovery, and IR-related processes. To achieve the above, RPs in PHOENI2X adopt the CACAO standard.

Considering the above, the utility security operators will need to define their own set of playbooks or tailor external playbooks received to their specific deployments, infrastructure, and preferred continuity, recovery, and overall IR strategies to accurately reflect their relevant processes. These can cover all five types of phases and requirements detailed in Section \ref{sec:Reqs}. An example playbook focusing on IR and business continuity for the AMI is presented in Fig. \ref{fig:playbookAMI}.

\section{Proof of concept - Automated Response to emulated attacks in an AMI testbed via CACAO Playbooks} \label{sec:caseStudy}
\subsection{Testbed Description}

\begin{figure}[bthp!]
     \centering
    \vspace{-1.2ex}
         \includegraphics[width=\columnwidth]{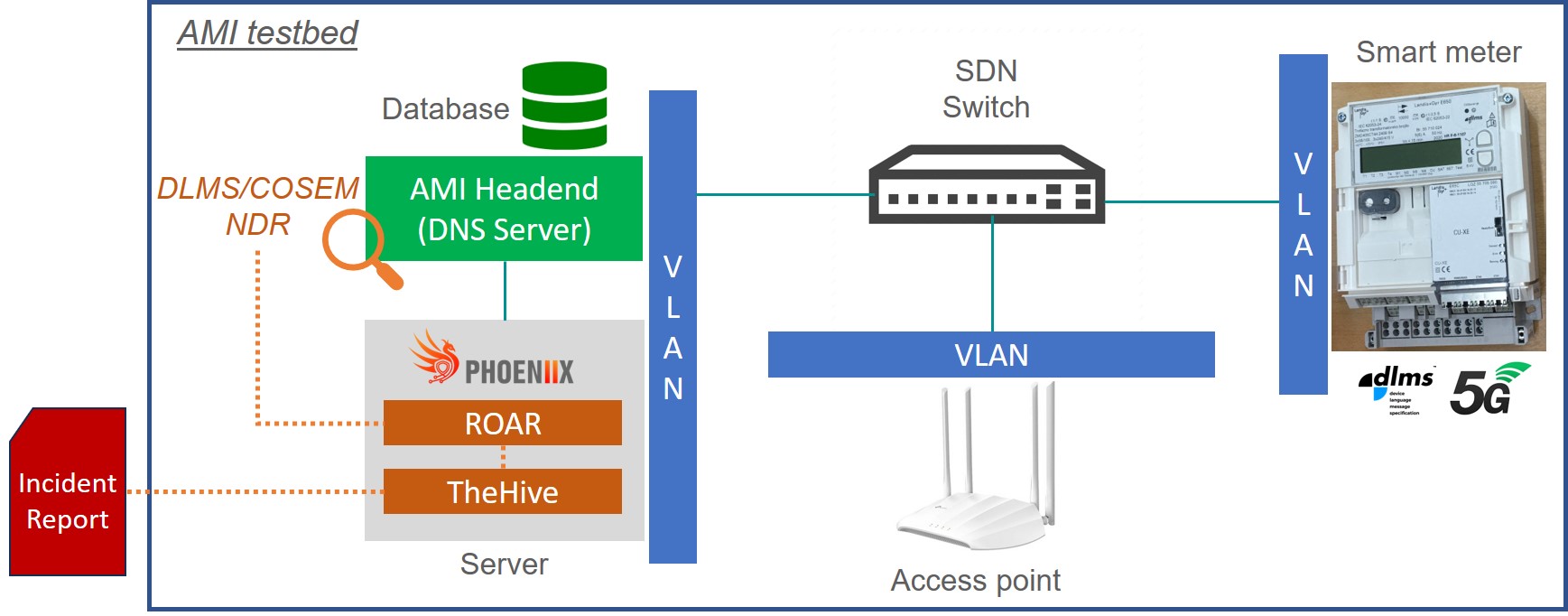}
     \caption{AMI testbed architecture}
     \label{fig:testbedArchitecture}
     \vspace{-3ex}
\end{figure}

The AMI testbed is a realistic emulation environment, including smart meters and an AMI Headend. The smart meters, manufactured by Landis+Gyr (E650 model compatible with DLMS/COSEM\footnote{https://www.landisgyr.eu/product/landisgyr-e650/}), are 5G-enabled, allowing measurement exchange with low latency. The AMI Headend used is based on the GuruX library\footnote{https://github.com/Gurux/Gurux.DLMS.UI.Net} and collects, aggregates, and monitors the exchanged network packets. The application logs from GuruX are stored in a PostgreSQL database. The smart meters and their deployment in the testbed architecture are illustrated in Fig. \ref{fig:testbedArchitecture}.

The figure shows that the testbed is split into different network segments (VLANs) to enforce an air gap between the critical systems (i.e., smart meters, AMI Headend, database with historical measurements) and the IT infrastructure (i.e., access point). Communication is handled through Aruba switches\footnote{https://www.arubanetworks.com/products/switches/} with Software Defined Networking (SDN) technologies \cite{nunes2014survey} for configuring and visualizing the network traffic flows. The infrastructure also includes containerized processes and applications (through docker images) using virtual environment computing nodes and a Type-1 hypervisor. Access to the infrastructure and the individual segments is restricted through a Virtual Private Network (VPN). The playbooks and associated tools for IR are deployed within a PHOENI2X server, which communicates with the AMI Headend and the database to retrieve the smart meter measurements.

The PHOENI2X server features the ROAR component of the framework, which supports the specification of RPs in a no-code, user-friendly manner and the execution and orchestration of the RPs. The execution and orchestration of multiple RPs in parallel is also supported, along with monitoring the progress/status of their execution and collecting any evidence generated. For the front-end specification and monitoring, the GUI of the Node-RED\footnote{https://nodered.org/} low-code programming tool for event-driven applications is leveraged, extending it with custom-made nodes and underlying execution capabilities to support the features of RPs. For case management, the server also features an instance of TheHive\footnote{https://thehive-project.org/}, an open-source security IR platform that facilitates incident case management. 


\subsection{Preparation of Cyber-Attack and Detection}

To launch an attack against the AMI system, an adversary may obtain access to the AMI Headend through supply chain attacks, phishing, leaked credentials, or even by exploiting a vulnerability to gain remote access to one or more smart meters. Nevertheless, such attacks can be challenging to carry out in practice, as smart meters are typically designed with built-in security features to prevent unauthorized access or tampering \cite{sari2020industrial}. Moreover, as the TCP session establishment in DLMS/COSEM for the exchange of smart meter measurements is initiated explicitly from the AMI Headend, DoS attacks \cite{yi2014denial} flooding the network with malicious or DLMS/COSEM service request packets \cite{sari2020industrial} are not possible on the Headend but only on the individual smart meters. Consequently, DoS attacks can cause processing overload on the smart meters.

Furthermore, an attack against multiple smart meters may significantly impact the utility in terms of financial losses or inaccurate historical measurements for forecasting load demand (as part of the energy production business process). Hence, a Distributed DoS (DDoS) attack against multiple smart meters was conducted within the testbed using 100 virtual emulated smart meters deployed in the PHOENI2X server. The emulated smart meters replicate the operational and network behavior of Landis+Gyr E650, which is used in industrial, commercial, and smart grid applications. 

Additionally, adversaries may leverage the communication between the smart meters and the AMI Headend, which is based on long-lasting TCP sessions \cite{tudor2015harnessing}. Address Resolution Protocol (ARP) spoofing techniques \cite{kulkarni2020mitm} can be used to perform a Man-in-The-Middle (MiTM) attack and intercept any existing security measure that is available in the smart meters. Specifically, DLMS/COSEM defines low (password) and high (encryption keys) security measures for data exchange during the TCP session establishment between the smart meter and the AMI Headend \cite{sari2020industrial}. 

\begin{figure}[bthp!]
\vspace{-3.1ex}
     \centering
         \includegraphics[scale=.805]{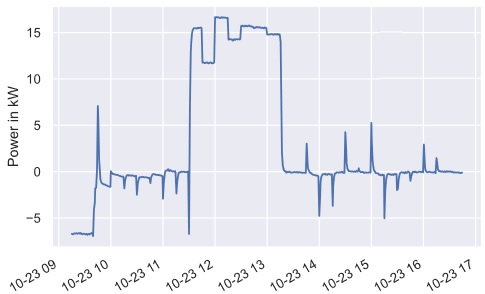}
     \caption{Emulated FDI attack in the apparent power}
     \label{fig:smartMeterattack}
     \vspace{-4ex}
\end{figure}

Ettercap \cite{pingle2018real} can be used to conduct MiTM and ARP spoofing attacks (Ettercap \cite{pingle2018real}) between the Landis+Gyr smart meter and the GuruX system (i.e., AMI Headend). Such attacks can be performed at a low-security level but are challenging at a high-security level. Another approach for initiating the attacks, irrespective of the security level, is obtaining leaked credentials to get direct access to the AMI Headend or the smart meter. As a next step, an adversary may perform FDI by altering the DLMS/COSEM Read Request/Responses or data tampering by dropping AMI measurements. An FDI attack can have a higher impact since false measurements lead to inaccurate predictions and billing. Hence, experiments were performed for both scenarios, but hereby, we focus on presenting the FDI and DDoS attacks on the virtually emulated smart meters. The apparent power measurements of the Alternate Current (AC) system upon the emulated FDI attack execution are illustrated in Fig. \ref{fig:smartMeterattack}. The figure depicts a spike in electricity usage/consumption as the attack starts at 11:30 am and ends at 01:15 pm. Furthermore, the negative values indicate an update in the AC waveform direction, affecting the apparent power. 

The cyber-attack detection was performed using a Network Detection and Response (NDR) tool that is based on the Zeek network security monitor \cite{haas2020zeek} and operates on industrial network protocols \cite{lekidis2022cyber}, such as DLMS/COSEM (see "DLMS/COSEM NDR" on Fig. \ref{fig:testbedArchitecture}). Specifically, the tool dissects the packets exchanged at the network level, interprets the DLMS/COSEM commands in them, and generates log files (i.e., XML or JSON format). The log files contain all the contextual protocol information (e.g., Association Request/Response, Read Request/Response \cite{sari2020industrial}). 

\begin{figure}[bthp!]
     \centering
     \vspace{-2ex}
         \includegraphics[scale=.65]{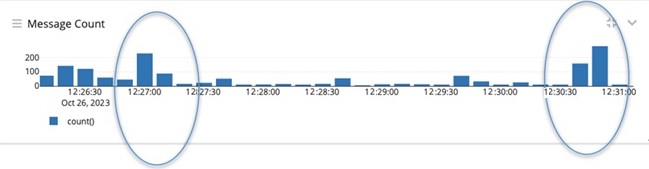}
     \caption{Smart meter DLMS/COSEM messages during the DDoS attack}
     \label{fig:DDoSattack}
     \vspace{-3ex}
\end{figure}

For detecting FDI attacks, the NDR tool builds a customized profile by clustering smart meters and their nominal values for the measurements according to a) the location where the smart meter is deployed (Section \ref{sec:background}), b) the area size where the consumption is measured (in $m^2$) and c) the seasonal/temperature variations. Deviations from the customized profile are considered anomalies and lead the NDR to generate an alert, which triggers further IR actions (see Section \ref{sec:mitigation}). On the other hand, the DDoS attack detection was based on increased DLMS/COSEM network messages exchanged from the smart meter(s), as depicted in Fig. \ref{fig:DDoSattack}.

\subsection{Mitigation, Remediation \& and Incident Reporting}
\label{sec:mitigation}

An important note for the following part of this paper is that even though we have implemented both DDoS and FDI attacks, we detail only the FDI response process due to space constraints. The FDI attack was considered more complicated to address, whereas best practices for DDoS prevention, such as load balancing approaches \cite{zebari2020distributed}, are present in literature, and hence, even though they were implemented, they are not detailed further. 

Upon the detection of the incident, the NDR generates an alert in the Micro Focus Common Event Format (CEF)\footnote{https://www.microfocus.com/documentation/arcsight/arcsight-smartconnectors-8.4/pdfdoc/cef-implementation-standard/cef-implementation-standard.pdf} containing all the necessary information about the cyber-attack, such that it can also be integrated with Security Information and Event Management (SIEM) tools, such as Wazuh\footnote{https://wazuh.com/}. Afterward, the NDR triggers the ROAR component to perform response actions. These actions initially include the modification of SDN switch network flows to reconfigure the segments of the AMI testbed. Hence, the infected hosts/systems will be isolated in a separate VLAN segment. To ensure an up-and-running and uninterrupted smart meter measurement exchange service and based on whether the AMI Headend is the system affected, a hot standby AMI Headend is activated. On the other hand, if a smart meter is compromised, a clean firmware installation is performed, or a patch is applied, considering an identified/detected vulnerability.

The response actions occur through dedicated POST requests in REST API calls, which include the network flow update. A fragment of an SDN switch API call for a network flow update is depicted in Listing \ref{lst:listing}, using anonymized user credentials in the authorization request. Specifically, the listing includes the creation of a new network flow for data transmission (\textit{"OUTPUT"} value for the \textit{"type"} parameter) and a dedicated port (\textit{"port"} parameter) in a specific network segment (\textit{"match"} parameter). The network flow is then included in a table with all flows, also receiving a unique entry number (\textit{"table\_id"} parameter). 
\begin{minipage}[bthp!]{.95\columnwidth}
\centering
\begin{lstlisting}[caption=SDN switch API for network flow update, label={lst:listing}, language=JSON, basicstyle=\small]
curl --location 'https://sdn-switch.com:10443/stats/flowentry/add'
--header 'Authorization: Basic user:ORWoIJZrgrb9S4jYUy0'
--data '{
    "dpid": 2876467493016320,
    "priority": 15,
    "actions": [{
            "type": 'OUTPUT',
            "port": 8
    }],
	"match": {
		"ipv4_src": '10.0.0.1',
		"eth_type": 2048
	},
    "table_id": 100
}'
\end{lstlisting}
\end{minipage}

\begin{figure}[bthp!]
    \includegraphics[width=\columnwidth]{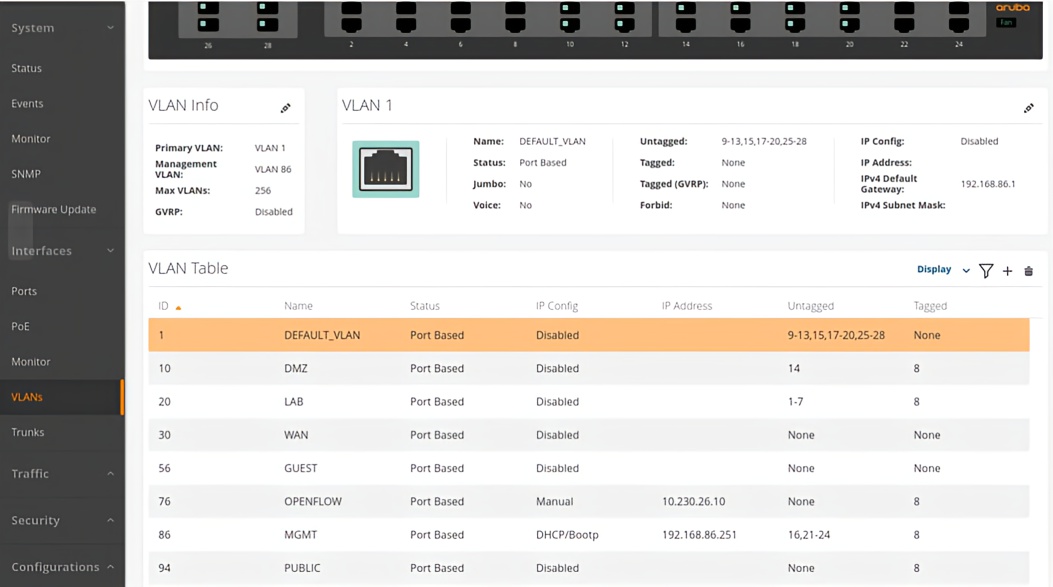}
    \caption{Infected host isolation in a VLAN through SDN switch}
    \label{fig:sdnController}
    \vspace{-3ex}
\end{figure}

The SDN switch ensures the isolated systems have zero interaction with the up-and-running AMI system until the attack is fully eradicated. The infected host isolation can be visualized through the Aruba SDN switch dashboard (Fig. \ref{fig:sdnController}).

Eradication occurs after the incident is mitigated within the sandboxing VLAN segment by installing software patches or clean firmware versions for the smart meters and the AMI Headend. Then, they are tested within the isolated network, and if they have no evidence of infection, the SDN switch API is again invoked to restore the isolated segment.

\begin{figure*}[bthp!]
     \centering
         \includegraphics[width=\textwidth, scale=.18]{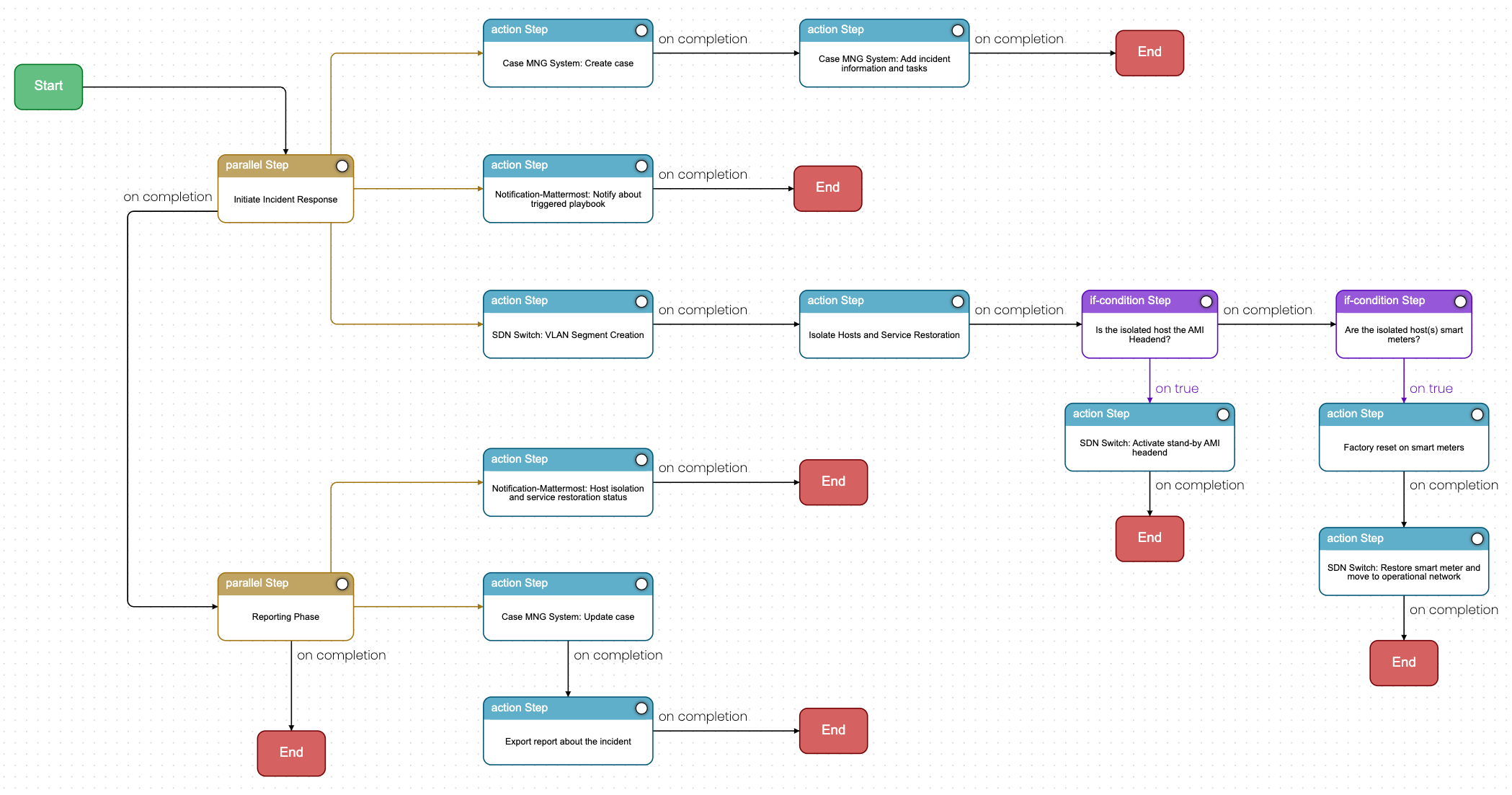}
     \caption{IR using a CACAO playbook executed by the ROAR component}
     \vspace{-5ex}
     \label{fig:playbookAMI}
\end{figure*}

Fig. \ref{fig:playbookAMI} illustrates the CACAO playbook that encodes the above actions, as visualized and eventually executed on the ROAR component of PHOENI2X (see "ROAR" in Fig. \ref{fig:testbedArchitecture}). Specifically, the CACAO playbook: 
\begin{compactenum}
    \item Is triggered from the DLMS/COSEM NDR detection mentioned above through a listener in ROAR.
    \item In parallel:
    \begin{compactenum}
        \item Opens a case on TheHive, also adding the relevant incident mitigation, remediation, and reporting tasks and passing over any relevant information (e.g., IPs of the victim and offending hosts, as captured by the NDR).
        \item Notifies the IR team about the case and the triggered playbook in real-time by sending a direct message (e.g., message via a messaging platform, like Slack
        or Mattermost
        , customizable depending on the tool of choice for the specific organization).
         \item Creates a new VLAN segment for isolating infected systems through the SDN switch.
    \end{compactenum} 
    \item Triggers host isolation and service restoration by directly interacting with the API of the SDN switch. Conditional logic is applied, meaning that: 
        \begin{compactenum}
        \item If the system exhibited abnormal behavior, is an AMI Headend, the suspicious Headend is placed in the sandbox VLAN segment, and an active standby Headend is deployed. 
        \item Otherwise, if it is a smart meter, it is placed in the new VLAN segment, and a firmware reinstallation is performed. Upon its completion, the smart meter is brought back to the operational network segment. 
        \item When both systems are suspicious (exhibited abnormal behavior), then both actions a) and b) are fully executed. 
        \end{compactenum}
    \item After containment/remediation are concluded and verified by the SDN switch, the automated playbook in parallel:
    \begin{compactenum}
        \item Updates the case on TheHive to document that all tasks have been successfully performed.
        \item Sends another direct message notification to the security operator to inform that the case has been handled (i.e., the playbook was successfully executed along with pivoting points to investigate further - for example, direct access to logs and details about the incident).
        \item Exports a detailed report that can facilitate the organizations' post-incident analysis and reporting requirements (each member state NCA/CSIRT provides templating approaches for reporting incidents). 
    \end{compactenum}
    \item Ends the playbook execution (recording all logs, as needed).
\end{compactenum}
Furthermore, considering the incident reporting and post-incident reporting requirements detailed in Section \ref{sec:Reqs}, reporting about (significant) incidents can occur in three distinct phases. Initially, when the incident is detected through the NDR and the IR team is informed, they should submit a warning within 24 hours to make the CSIRT, or where applicable, the competent authority, aware of the incident, allowing also the OES to seek assistance whenever necessary. Then, the warning is followed by a notification to the CSIRT containing an initial incident assessment and Indicators of Compromise (IoCs) within 72 hours of the incident occurrence. Afterward, a final report is prepared no later than one month after the incident occurrence, including all the incident details and the mitigation actions that took place. However, the automated approach presented in this paper demonstrated a rapid response against specific types of incidents, including playbook nodes for assisting with incident reporting, allowing the generation and dissemination of one comprehensive report in a very timely manner. In addition, we can transmit and exchange relevant CTI and playbook templates that are considered highly reliable to address particular types of attacks and threats in support of European cross-border cyber security cooperation and collaboration.


\vspace{-3mm}
\section{Discussion} \label{sec:discussion}


\subsection{Evaluation Metrics for the Proposed Method}

The effectiveness of the proposed approach was measured via established IR metrics \cite{torkura2019slingshot}, such as the Mean Time To Respond (MTTR). MTTR is the average time for an incident to be resolved completely and, in the context of this research, after it has been detected. 
Overall, lower MTTR is considered better to ensure the effectiveness and efficiency of an IR plan. In our scenario, which was motivated by a real-world use case, the utility did not have any incident response automation capability before deploying the PHOENI2X toolset. Hence, the detection and response to cyber-attacks were based on manual operator investigations on NDR alerts and logs from firewalls and the Virtual Private Network (VPN), mechanisms that were also employed in our testbed infrastructure. Moreover, these investigations only occurred when abnormal behavior in the AMI business service was observed through the NDR. In particular, we witnessed that upon a successful FDI attack, the operators responsible for security operations did not notice that an incident had occurred until they manually performed triage and investigated the alerts from the NDR. Therefore, the proposed IR method improves the current process efficiency (while remaining fault-tolerant) through the ROAR orchestration of the entire detection, response, and post-incident procedural flow.

In particular, through the introduced automation, we measured a reduction of 96\% in the MTTR in the FDI attack scenario and 98\% in the DDoS attack scenario, taking overall the response time for these two attack types from a few hours to a few minutes. Specifically, all the steps in the CACAO playbook are completed within 5-10s, except the reinstallation of firmware, which is the most time-consuming process. This process is split into three steps: 1) obtaining the firmware that needs to be installed, 2) installing the firmware on the smart meter, and 3) rebooting the device. Even though the first step usually takes 1 minute at most, the second and third steps may take 15 minutes each. The reason is that during the firmware installation, all the configurations, processes, and settings have to be added, and upon booting, the smart meter has to initiate all the processes and the DLMS/COSEM parameters set during the firmware installation.




The DDoS attack had a lower MTTR than the FDI due to the more straightforward IR process that was instantiated, such as load balancing and adding firewall rules. This resulted in a faster playbook execution even though the number of smart meters where the mitigation measures were applied was higher.  

Concerning impact, it is worth noting that the previous day's smart meter measurements are used as historical data in the day-ahead forecasting/planning algorithms. Specifically, suppose the forecasting processes produce inaccurate energy demand predictions for the day-ahead market. In that case, this will lead to an imbalanced production/demand equilibrium, which in turn causes discrepancies in the grid operation. Additionally, when the billing/clearance period for the customers is reached (i.e., 1-3 months), the utility might suffer substantial financial/legal penalties if the smart meter measurements were altered and resulted in wrong billing. 

Finally, apart from the day-ahead forecasting and billing processes, another process might also be indirectly affected by the absence of an IR method, i.e., the EV charging business. In particular, EV charging is also based on smart meter measurements for the energy delivered to EVs and requires a significantly reduced Response Time Objective (RTO) as well as a Maximum Tolerable Period of Disruption (MTPD) when considering the Business Impact Analysis based on ISO/TS 22317:2021 standard \cite{ISO22317} guidelines. Hence, the absence of a highly automatable IR process may cause severe financial, reputational, and regulatory consequences.

\subsection{Ongoing Work and Future Directions}

Ongoing work focuses on conducting large-scale cyber-attacks, including different types of attacks on the AMI system, identifying potential cascading effects and impacts on the business continuity of further services (such as energy demand forecasting), and incorporating automation for their detection and resolution. To this end, less likelihood but with significantly higher impact attacks for AMI systems will be considered, such as malware spreading, where attackers employ common malware such as Mirai to infect simultaneously multiple smart meters, including the AMI Headend, to manipulate their operational status, and create a botnet controlled through a Command \& Control server. Furthermore, stealthy malware targeting Linux environments may also be employed like in the case of Shikitega\footnote{https://cybersecurity.att.com/blogs/labs-research/shikitega-new-stealthy-malware-targeting-linux}. 

Finally, our testbed is emulating an AMI infrastructure. Nevertheless, smart meters are also included inside EV chargers to measure and bill customers based on the power delivered to the EVs. Additionally, attacks in EV charger meters can propagate toward the EVs, causing cascading damages that may jeopardize driver safety. Such cascading damages are critical, and since e-mobility is currently an emerging utility business, an orchestrated and automated IR process like the one presented in the paper is deemed extremely necessary for security and safety. 


\section{Conclusions \& next steps} \label{sec:conclusion}

This paper presented a novel method for orchestrating and automating the cyber security of AMI systems and ensuring business continuity against specific attack scenarios. In particular, we focused on ensuring continuous availability of the smart meter data exchange service while a threat is contained and remediated. Automation is accomplished through CACAO playbooks, which orchestrate courses of action and actuate specific systems and components in the context of a response plan. The method is applied on an AMI testbed where MiTM/ARP spoofing and supply chain/leaked credentials are used to access the testbed and emulate FDI and data tampering attacks to alter or drop smart meter measurements. Additionally, DDoS attacks were performed on emulated smart meters to achieve network flooding with malicious/service request packets, which caused smart meter processing overload. The attacks are detected and mitigated using the PHOENI2X tools, and an incident report is generated to comply with incident reporting requirements, as imposed by EU Member States' national legislation and NIS2. The proposed method reduced the MTTR to FDI and DDoS attacks by 96\% and 98\%, respectively.


%
\IEEEpeerreviewmaketitle


\section*{Acknowledgment}
This work has received funding from the European Union’s Horizon Europe programme under Grant Agreement No. 101070586 (PHOENI2X project).

\bibliographystyle{IEEEtran}
\bibliography{biblio}

\end{document}